\documentclass[3p,times,procedia]{elsarticle}
\usepackage{nupha_ecrc}

\usepackage{wrapfig}

\volume{00}

\firstpage{1}

\journalname{Nuclear Physics A}

\runauth{}

\jid{nupha}

\jnltitlelogo{Nuclear Physics A}

\usepackage{amssymb}

\usepackage[figuresright]{rotating}

\newcommand{\xt}{{\mathbf{x}_T}}

\newcommand{\yt}{{\mathbf{y}_T}}

\newcommand{\nc}{{N_\mathrm{c}}}

\newcommand{\tr}{\, \mathrm{Tr} \, }

\newcommand{\qs}{Q_\mathrm{s}}

\newcommand{\fig}{Fig.~}

\newcommand{\nr}[1]{(\ref{#1})}

\begin{document}

\begin{frontmatter}




\dochead{XXVIIth International Conference on Ultrarelativistic Nucleus-Nucleus Collisions\\ (Quark Matter 2018)}

\title{
Forward rapidity isolated photon production in  proton-nucleus collisions
}


\author[cea]{B. Duclou\'e}
\author[jyu,hip]{T. Lappi$^{*,}$\corref{cor}}
\author[jyu]{H. M\"antysaari}

\cortext[cor]{Speaker}

\address[cea]{
Institut de Physique Th{\'e}orique, Universit{\'e} Paris Saclay, CEA, CNRS, F-91191 Gif-sur-Yvette, France%
}

\address[jyu]{Department of Physics, University of Jyv\"askyl\"a %
 P.O. Box 35, 40014 University of Jyv\"askyl\"a, Finland%
}
\address[hip]{
Helsinki Institute of Physics, P.O. Box 64, 00014 University of Helsinki,
Finland%
}

\begin{abstract}
We calculate isolated photon production at forward rapidities in proton-nucleus collisions in the Color Glass Condensate framework. Our calculation uses dipole cross sections solved from the running coupling Balitsky-Kovchegov equation with an initial condition fit to deep inelastic scattering data and extended to nuclei with an optical Glauber procedure that introduces no additional parameters beyond the basic nuclear geometry. We present predictions for future forward RHIC and LHC measurements. The predictions are also compared to updated results for the nuclear modification factors for pion production, Drell-Yan dileptons and $J/\psi$ mesons in the same forward kinematics, consistently calculated in the same theoretical framework. We find that leading order, running coupling high energy evolution in the CGC picture leads to a significant nuclear suppression at forward rapidities. This nuclear suppression is stronger for photons than for pions. We also discuss how this might change with next-to-leading order high energy evolution.
\end{abstract}


\end{frontmatter}


\section{Introduction}
\label{sec:intro}

The initial stage of a heavy ion collision is dominated by the color field of small-$x$ gluons in the colliding nuclei. One would like to independently probe this color field with a simple dilute probe. The cleanest way to do this experimentally would be in deep inelastic scattering measurements. An alternative is to study particle production in pA- or pp-collisions at forward rapidity, which we will discuss here. In particular we discuss signals of gluon saturation, and address the effects of small-$x$  renormalization group evolution.

The discussion here is based on several recent works. We will first discuss the most recent one, the calculation of isolated photon production in proton-nucleus collisions in Ref.~\cite{Ducloue:2017kkq}. We will then compare the nuclear suppression factors to those for other prcoesses calculated in the same framework: $J/\psi$ and single inclusive hadron production. Finally we speculate about what could change when going from these leading order calculations to next-to-leading order in perturbation theory. 

All of the processes studied here can be understood in terms of an eikonal picture. Here the scattering process depends on the light-like Wilson line of the color field of the target nucleus, which is the eikonal scattering amplitude of the colored probe. In fact gauge invariant cross sections depend on color singlet  operators made out of these Wilson lines, the most common being the color dipole
\begin{equation} \label{eq:dip}
N \left(|\xt-\yt|\right) = 1-\left< \frac{1}{\nc}\tr V^\dag(\xt)V(\yt)\right>.
\end{equation}
For small dipoles $N(r) \to 0$, a phenomenon known as color transparency. For large dipoles, on the other hand, $N(r)\to 1$. Thus somewhere in between there is a typical correlation length of the Wilson lines, denoted by $1/\qs$. The corresponding momentum scale $\qs$, marking the transition from a dilute high momentum regime to one dominated by nonlinearities, is known as the saturation scale. 

Here we obtain $N(r)$  from the 
 ``MV$^e$'' parametrization of Ref.~\cite{Lappi:2013zma}. Here the dipole amplitude for a proton target at the initial momentum fraction $x_0=0.01$ is parametrized in terms of 3 parameters. It is then evolved to lower momentum fractions $x$ using a leading order running coupling Balitsky-Kovchegov (BK) evolution equation. 
 The free parameters are determined by inclusive HERA DIS cross sections and  extended from proton to nuclear targets using an optical Glauber framework. Thus no additional parameters apart from standard Woods-Saxon geometry are required to describe nuclei.

\section{Photon production at forward rapidity}

\begin{figure}
\includegraphics[width=0.45\textwidth]{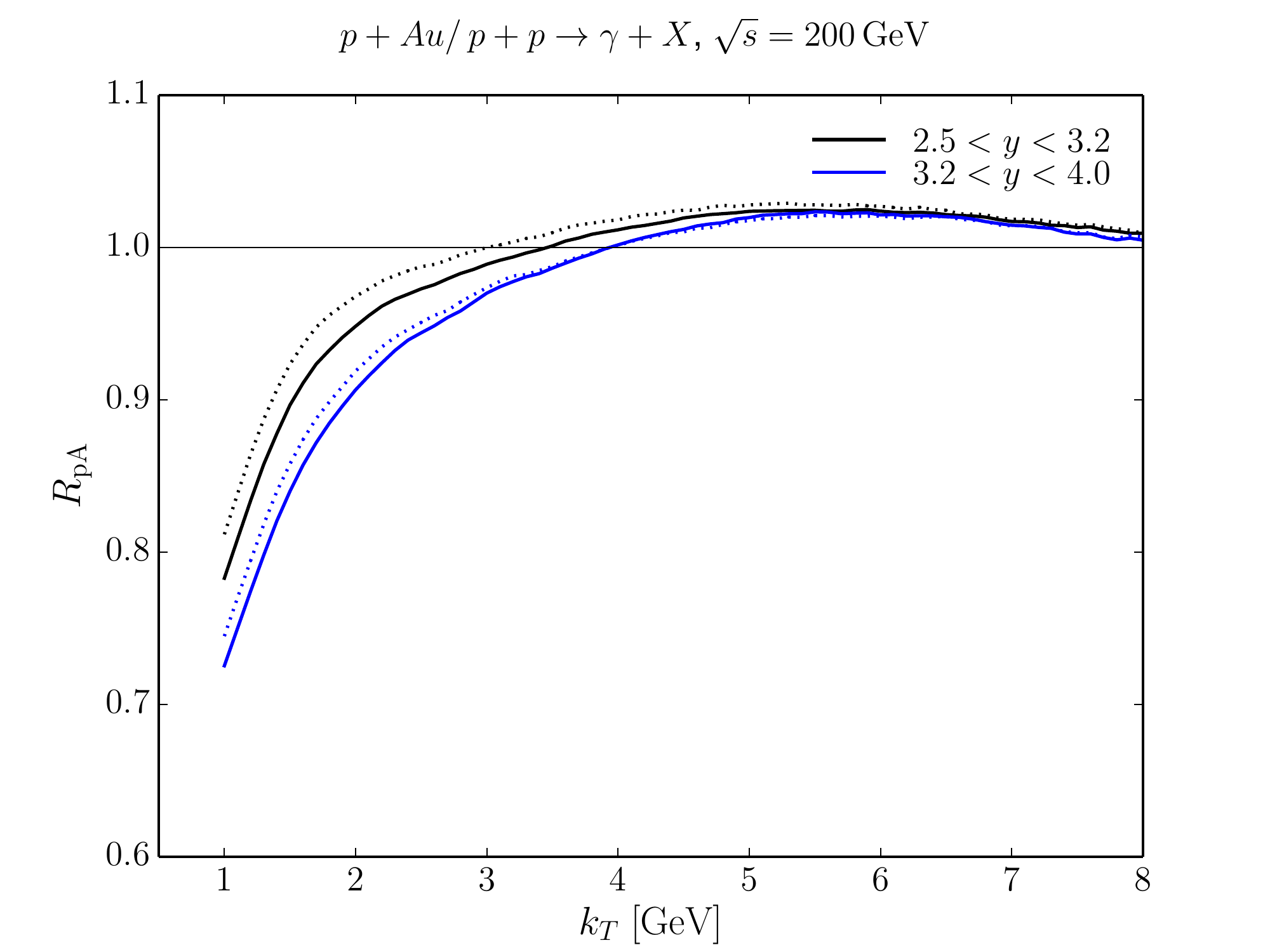}
\hfill
\includegraphics[width=0.45\textwidth]{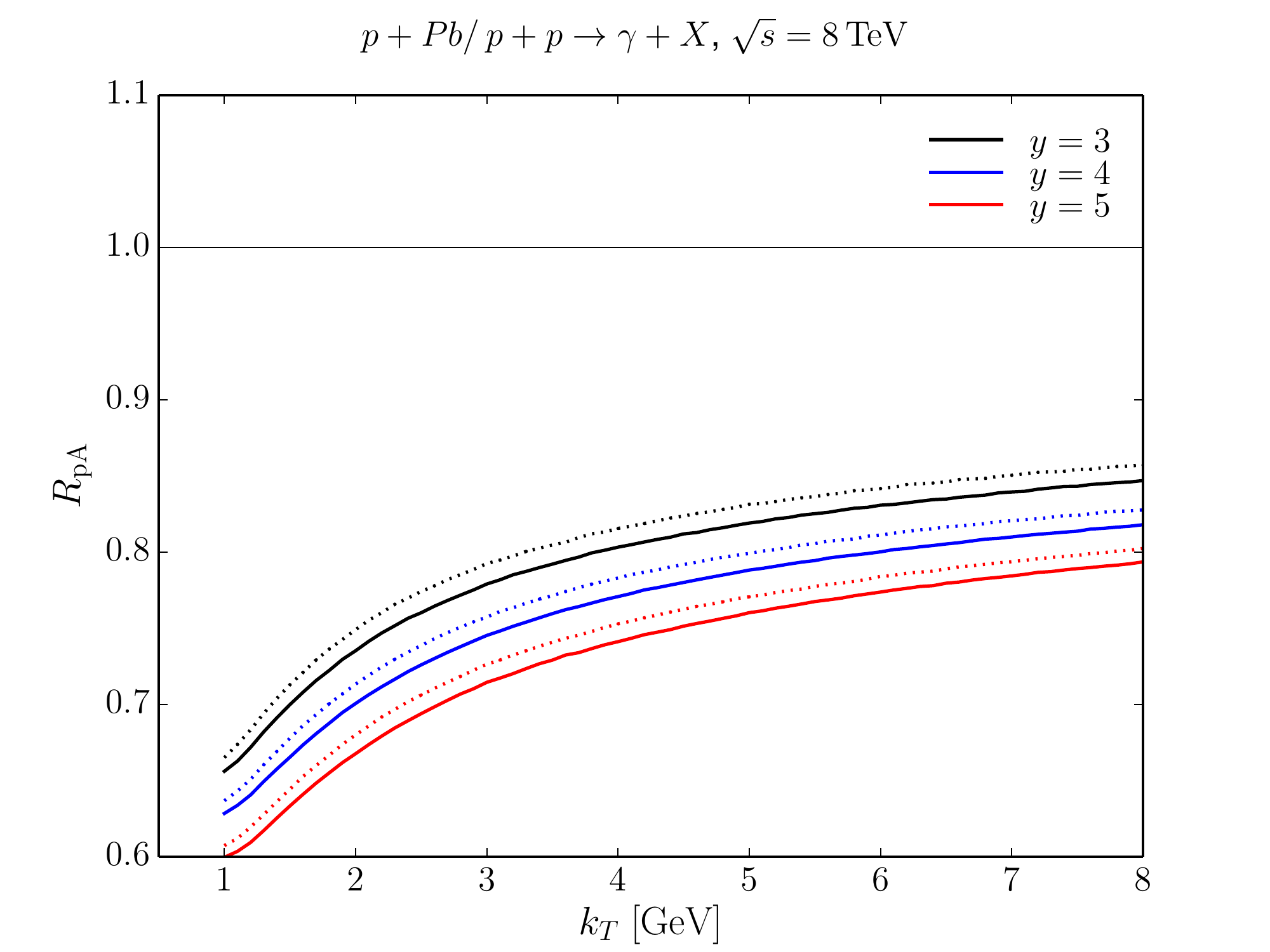}
\caption{Nuclear modification factor $R_{pA}$ for isolated photon production. Left: RHIC energy $\sqrt{s}=200$GeV, Right: LHC energy $\sqrt{s}=8$TeV. The isolation cuts are $R=0.4$ (solid line) and $R=0.1$ (dashed line) Plots from \cite{Ducloue:2017kkq}.}
\label{fig:photonrpa}
\end{figure}

In this framework photons are produced in the following way. A quark (which at forward rapidity comes from the high-$x$ part of the probe and can thus be described by conventional collinear parton distributions) passes through the color field of the target nucleus, picking up a Wilson line factor. Either before or after the collision, it can emit a real or virtual photon. The cross section for this process, expressed in terms of the same dipole amplitude~\nr{eq:dip} that we fit to DIS data was derived in Ref.~\cite{Gelis:2002ki}. It contains additional divergences when the photon becomes collinear with the outgoing quark, which we regulated with an isolation cut $ \sqrt{(y_\gamma - y_q)^2 + (\phi_{\gamma}-\phi_q)^2}>R$.

The result for the nuclear modification factor is shown in \fig\ref{fig:photonrpa} for both RHIC and LHC energies. In these results we can distinguish two effects. Firstly, already at RHIC energies photon production is suppressed at transverse momenta that are small compared to the saturation scale of the target. This suppression can be undestood as a clear indication of gluon saturation; the existence of a characteristic transverse momentum scale that is higher for nuclei than for protons. Secondly, when going to LHC energies one moves to much smaller values of the momentum fraction $x$. In this regime the nuclear suppression extends to much higher values of the photon transverse momentum. This is an effect of the high energy (BK) evolution which leads to a modification known as \emph{geometric scaling} in the gluon distribution of the target; we will discuss this effect further below.

\section{Comparisons with other dilute probes}

\begin{figure}
\hfill
\includegraphics[height=3.8cm]{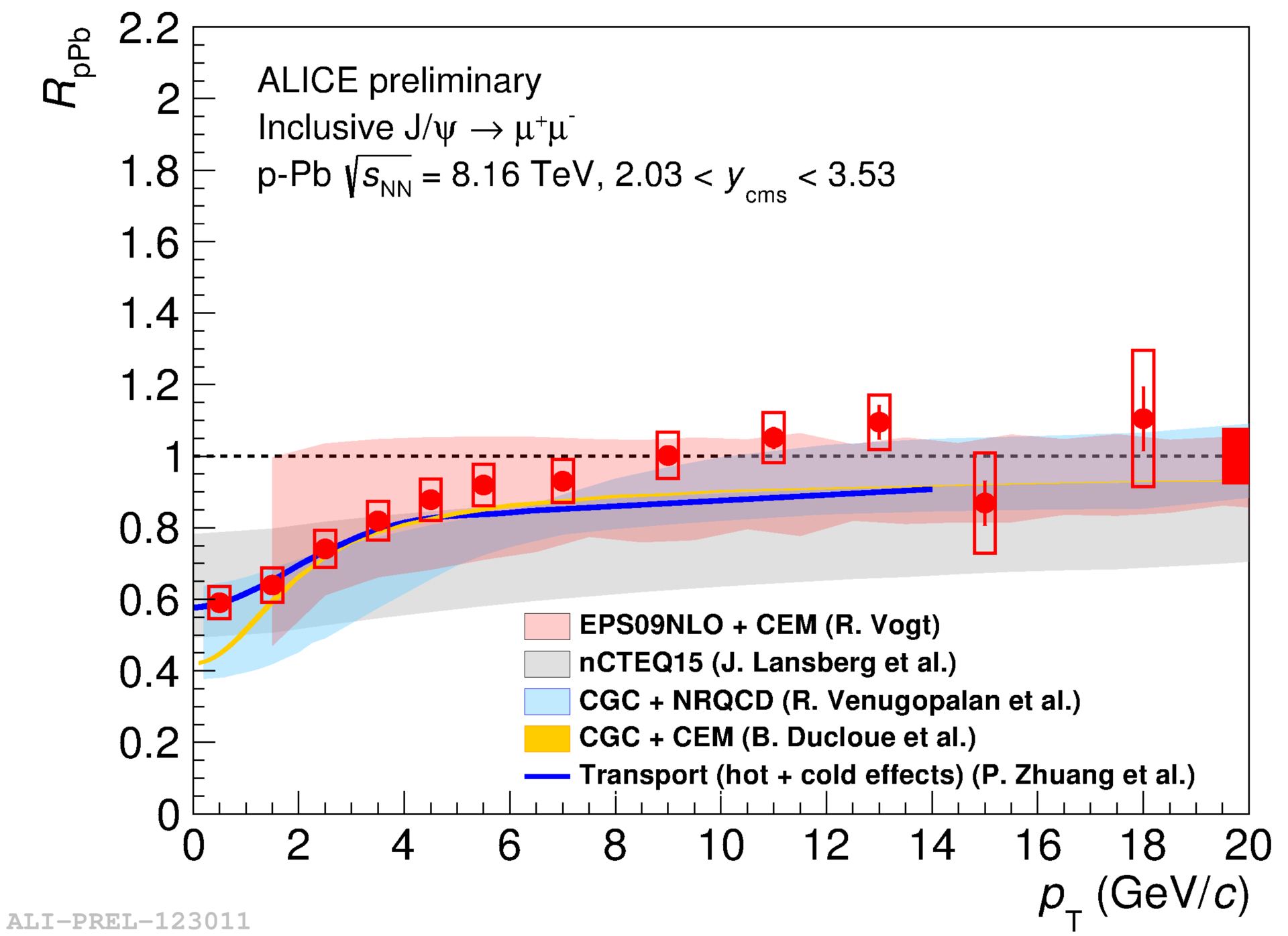}
\hfill
\includegraphics[height=3.8cm]{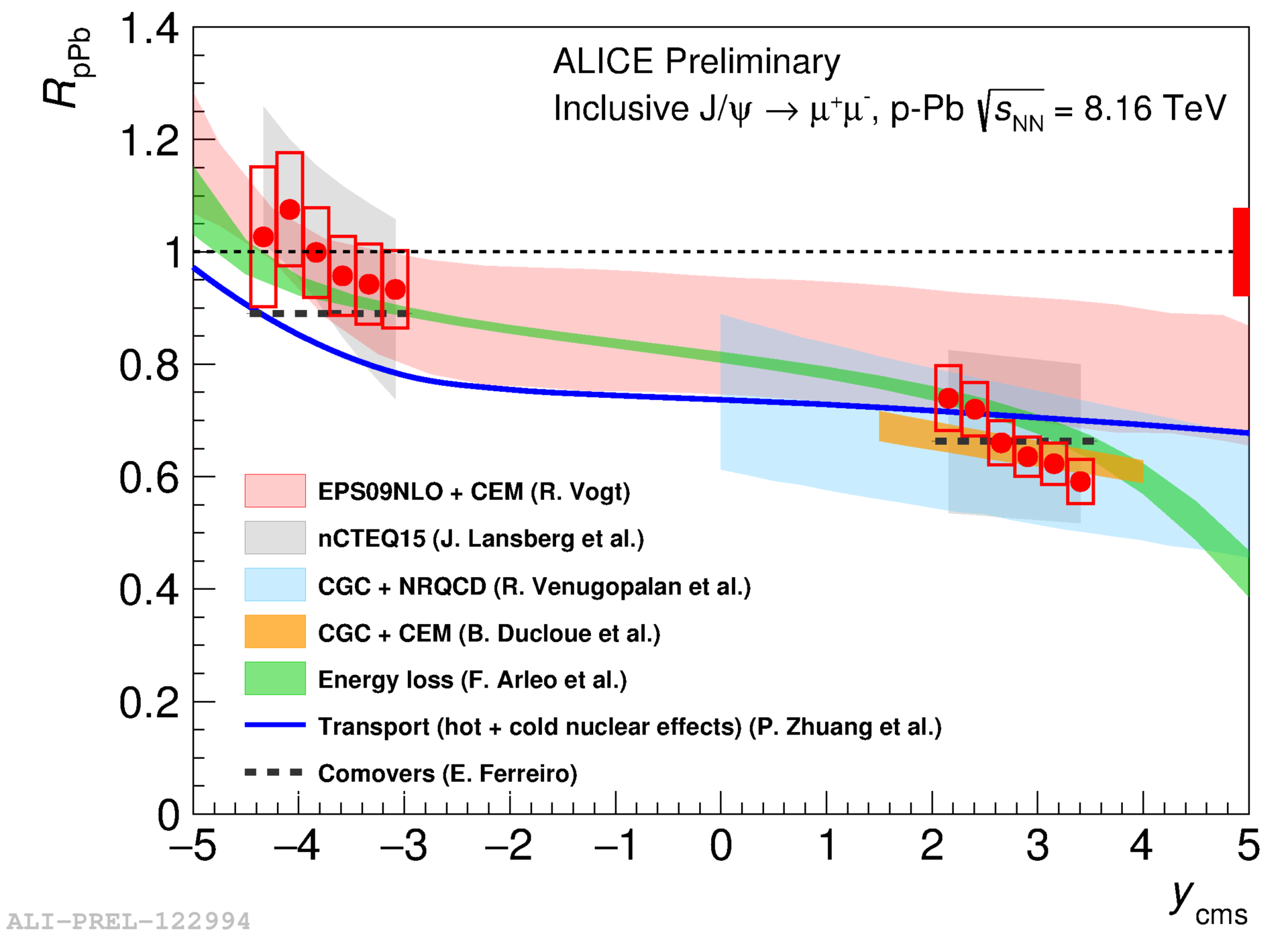}
\hfill \rule{1pt}{0pt}
\caption{ALICE results for $J/\psi$ \cite{Tarhini:2017xtb}. The theory calculation discussed in this paper is the CGC+CEM one from \cite{Ducloue:2015gfa,Ducloue:2016pqr}}
\label{fig:jpsi}
\end{figure}

\begin{figure}
\hfill
\includegraphics[height=3.5cm]{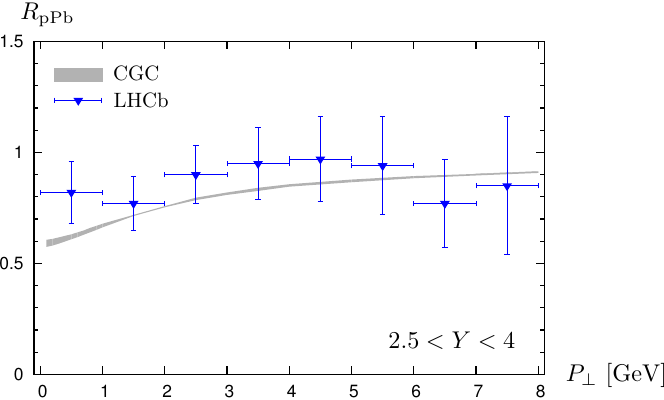}
\hfill 
\includegraphics[height=3.5cm]{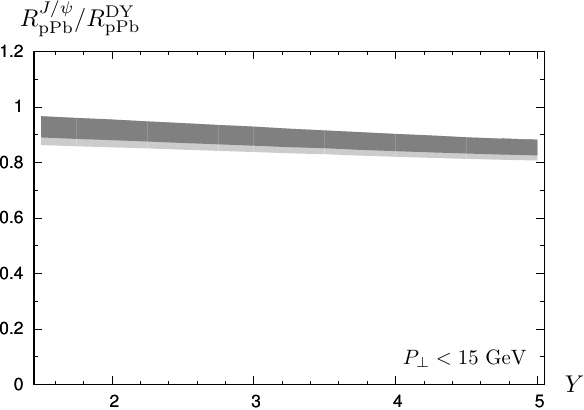}
\hfill \rule{1pt}{0pt}
\caption{Nuclear modifiction factor for $D$-mesons \cite{Ducloue:2016ywt} (left) and double ratio of nuclear modification factor for $J/\psi$ divided by the one for Drell-Yan-dileptons~\cite{Ducloue:2017zfd} (right). Note that the LHCb data in this plot is preliminary and the normalization in the finalized data~\cite{Aaij:2017gcy} is somewhat lower.}
\label{fig:DDY}
\end{figure}
One can also calculate other cross sections using exactly the same parametrization of the target color field. Figure~\ref{fig:jpsi} shows the nuclear modification ratio for $J/\psi$ mesons as a function of rapidity and transverse momentum.  One again distinguishes the same features as for photons: a suppression at low transverse momenta from gluon saturation, and an extension of this suppression to higher transverse momenta at forward rapidity at the LHC   due to  high energy evolution. The same features can also be seen in $D$-meson production and Drell-Yan cross sections in \fig\ref{fig:DDY}; this is a conceptually important comparison since in cold nuclear matter energy loss models the latter would behave in a very different way~\cite{Arleo:2015qiv}.

\begin{wrapfigure}{R}{0.35\textwidth}
\centerline{\includegraphics[height=4cm]{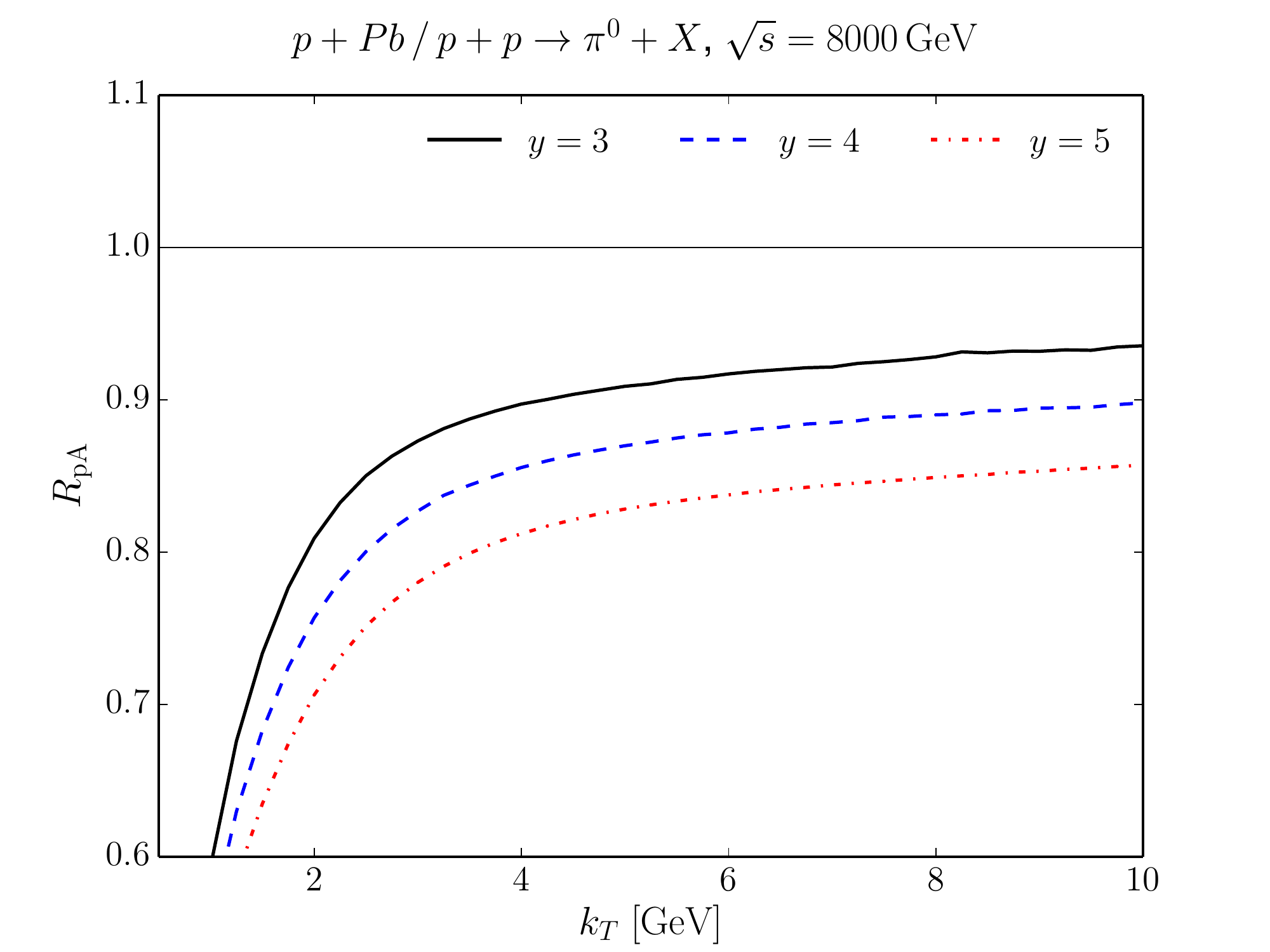}}
\caption{Nuclear modification factor for pions at LHC energies, \cite{Ducloue:2017kkq}}
\label{fig:light} 
\end{wrapfigure}
For light hadrons the story is almost the same. Figure~\ref{fig:light}   shows  the  pion $R_{pA}$ at  the same LHC kinematics as for photons in \fig\ref{fig:photonrpa}. The suppression for pions is not as large. The most likely explanation for this effect lies in the different kinematics of these two processes, when calculated in the CGC framework. Light hadron production is here calculated from the production of single quarks (or gluons, to a lesser extent) by scattering off the target color field, using collinear fragmentation functions. Photon production, on the other hand, involves  the production of a second unmeasured particle, the quark. Thus  for pions, a high momentum pion always requires high momentum gluons from the target, i.e. in a region where saturation is not important and there is little nuclear suppression. For photon production, on the other hand, it is possible to produce a high momentum photon from a small momentum exchange with the target, with the additional quark taking the recoil. This contribution is suppressed in nuclei, and thus photon production has a smaller $R_{pA}$ at higher momenta.  At next-to-leading order also light hadron production involves final state radiation, and the kinematics resembles the case of photons (see e.g. the recent work~\cite{Ducloue:2017dit} also reported in this conference). One could expect this to have an effect also on the nuclear suppression $R_{pA}$.

Let us finally speculate on the large nuclear suppression observed even at high momenta (even $\gg \qs$) at forward LHC kinematics. This has since quite a long time~\cite{Albacete:2003iq,Kharzeev:2003wz} been understood as a consequence of ``geometric scaling''. This term refers to a feature of high energy evolution  that changes the ``anomalous dimension'', i.e. the power $\gamma$ in the small-distance behavior of the dipole amplitude~\nr{eq:dip} $N(r) \sim r^{2\gamma}$ from $\gamma=1$ in the initial condition  to $\gamma\approx 0.6 \dots 0.8$. Unfortunately we do not have a full NLO calculation of the nuclear modification factor for any of the processes discussed here, although constant progress on this front is being made. However, it has been noticed~\cite{Lappi:2016fmu} that collinearly resummed NLO BK evolution practically freezes the anomalous dimension to its initial value. This could hint at a strong possible effect of NLO corrections even on $R_{pA}$, where most of the expected NLO corrections cancel in the ratio.

\section{Conclusions}

We have by now quite a large set of predictions for forward $pA$  cross sections in  a consistent framework: light hadrons using  fragmentation functions, real, virtual photons and  heavy quarks. In all of these BK evolution leads intrinsically to a forward rapidity  nuclear suppression. However, this conclusion comes with the caveat that the calculations so far are at leading order. In contrast to NLO, this leads to different 
kinematics for light parton ($q,g$)  vs. $Q\bar{Q},\gamma, \gamma^*$ processes. Also one expects NLO effects in the BK evolution to slow down the forward suppression. For a bigger picture understanding of the effects of gluon saturation it is important to simultaneously look at more exclusive observables, such as  multiparticle correlations that are also discussed in this conference~\cite{Albacete:2018ruq}.

\paragraph{Acknowledgements}
This work has been supported by the Academy of Finland, projects 267321 and 303756, and by the European Research Council, grant ERC-2015-CoG-681707. The work of B.D. is supported by the Agence Nationale de la Recherche project ANR-16-CE31-0019-01.





\bibliographystyle{elsarticle-num}
\bibliography{spires}

\begin{thebibliography}{10}
\expandafter\ifx\csname url\endcsname\relax
  \def\url#1{\texttt{#1}}\fi
\expandafter\ifx\csname urlprefix\endcsname\relax\def\urlprefix{URL }\fi
\expandafter\ifx\csname href\endcsname\relax
  \def\href#1#2{#2} \def\path#1{#1}\fi

\bibitem{Ducloue:2017kkq}
B.~Duclou{\'e}, T.~Lappi, H.~M{\"a}ntysaari, Isolated photon production in
  proton-nucleus collisions at forward rapidity, Phys. Rev. D97~(5) (2018)
  054023.
\newblock \href {http://arxiv.org/abs/1710.02206} {\path{arXiv:1710.02206}},
  \href {http://dx.doi.org/10.1103/PhysRevD.97.054023}
  {\path{doi:10.1103/PhysRevD.97.054023}}.

\bibitem{Lappi:2013zma}
T.~Lappi, H.~M{\"a}ntysaari, Single inclusive particle production at high
  energy from {HERA} data to proton-nucleus collisions, Phys. Rev. D88 (2013)
  114020.
\newblock \href {http://arxiv.org/abs/1309.6963} {\path{arXiv:1309.6963}},
  \href {http://dx.doi.org/10.1103/PhysRevD.88.114020}
  {\path{doi:10.1103/PhysRevD.88.114020}}.

\bibitem{Gelis:2002ki}
F.~Gelis, J.~Jalilian-Marian, Photon production in high-energy proton nucleus
  collisions, Phys. Rev. D66 (2002) 014021.
\newblock \href {http://arxiv.org/abs/hep-ph/0205037}
  {\path{arXiv:hep-ph/0205037}}, \href
  {http://dx.doi.org/10.1103/PhysRevD.66.014021}
  {\path{doi:10.1103/PhysRevD.66.014021}}.

\bibitem{Tarhini:2017xtb}
M.~Tarhini, Charmonium production in {Pb-Pb} and {p-Pb} collisions at forward
  rapidity measured with {ALICE}, Nucl. Phys. A967 (2017) 588--591.
\newblock \href {http://dx.doi.org/10.1016/j.nuclphysa.2017.05.027}
  {\path{doi:10.1016/j.nuclphysa.2017.05.027}}.

\bibitem{Ducloue:2015gfa}
B.~Duclou{\'e}, T.~Lappi, H.~M{\"a}ntysaari, Forward {$J/\psi$} production in
  proton-nucleus collisions at high energy, Phys. Rev. D91 (2015) 114005.
\newblock \href {http://arxiv.org/abs/1503.02789} {\path{arXiv:1503.02789}},
  \href {http://dx.doi.org/10.1103/PhysRevD.91.114005}
  {\path{doi:10.1103/PhysRevD.91.114005}}.

\bibitem{Ducloue:2016pqr}
B.~Duclou{\'e}, T.~Lappi, H.~M{\"a}ntysaari, Forward {$J/\psi$} production at
  high energy: centrality dependence and mean transverse momentum, Phys. Rev.
  D94~(7) (2016) 074031.
\newblock \href {http://arxiv.org/abs/1605.05680} {\path{arXiv:1605.05680}},
  \href {http://dx.doi.org/10.1103/PhysRevD.94.074031}
  {\path{doi:10.1103/PhysRevD.94.074031}}.

\bibitem{Ducloue:2016ywt}
B.~Duclou{\'e}, T.~Lappi, H.~M{\"a}ntysaari, Forward {$J/\psi$} and {$D$} meson
  nuclear suppression at the {LHC}, Nucl. Part. Phys. Proc. 289-290 (2017)
  309--312.
\newblock \href {http://arxiv.org/abs/1612.04585} {\path{arXiv:1612.04585}},
  \href {http://dx.doi.org/10.1016/j.nuclphysbps.2017.05.071}
  {\path{doi:10.1016/j.nuclphysbps.2017.05.071}}.

\bibitem{Ducloue:2017zfd}
B.~Duclou{\'e}, Nuclear modification of forward {Drell-Yan} production at the
  {LHC}, Phys. Rev. D96~(9) (2017) 094014.
\newblock \href {http://arxiv.org/abs/1701.08730} {\path{arXiv:1701.08730}},
  \href {http://dx.doi.org/10.1103/PhysRevD.96.094014}
  {\path{doi:10.1103/PhysRevD.96.094014}}.

\bibitem{Aaij:2017gcy}
R.~Aaij, et~al., Study of prompt {D$^{0}$} meson production in {$p$Pb}
  collisions at {$ \sqrt{s_{\mathrm{NN}}}=5 $ TeV}, JHEP 10 (2017) 090.
\newblock \href {http://arxiv.org/abs/1707.02750} {\path{arXiv:1707.02750}},
  \href {http://dx.doi.org/10.1007/JHEP10(2017)090}
  {\path{doi:10.1007/JHEP10(2017)090}}.

\bibitem{Arleo:2015qiv}
F.~Arleo, S.~Peign{\'e}, Disentangling shadowing from coherent energy loss
  using the {Drell-Yan} process, Phys. Rev. D95~(1) (2017) 011502.
\newblock \href {http://arxiv.org/abs/1512.01794} {\path{arXiv:1512.01794}},
  \href {http://dx.doi.org/10.1103/PhysRevD.95.011502}
  {\path{doi:10.1103/PhysRevD.95.011502}}.

\bibitem{Ducloue:2017dit}
B.~Duclou{\'e}, E.~Iancu, T.~Lappi, A.~H. Mueller, G.~Soyez, D.~N.
  Triantafyllopoulos, Y.~Zhu, Use of a running coupling in the {NLO}
  calculation of forward hadron production, Phys. Rev. D97~(5) (2018) 054020.
\newblock \href {http://arxiv.org/abs/1712.07480} {\path{arXiv:1712.07480}},
  \href {http://dx.doi.org/10.1103/PhysRevD.97.054020}
  {\path{doi:10.1103/PhysRevD.97.054020}}.

\bibitem{Albacete:2003iq}
J.~L. Albacete, N.~Armesto, A.~Kovner, C.~A. Salgado, U.~A. Wiedemann, Energy
  dependence of the {Cronin} effect from non-linear {QCD} evolution, Phys. Rev.
  Lett. 92 (2004) 082001.
\newblock \href {http://arxiv.org/abs/hep-ph/0307179}
  {\path{arXiv:hep-ph/0307179}}, \href
  {http://dx.doi.org/10.1103/PhysRevLett.92.082001}
  {\path{doi:10.1103/PhysRevLett.92.082001}}.

\bibitem{Kharzeev:2003wz}
D.~Kharzeev, Y.~V. Kovchegov, K.~Tuchin, {Cronin} effect and {high-p(T)}
  suppression in {p A} collisions, Phys. Rev. D68 (2003) 094013.
\newblock \href {http://arxiv.org/abs/hep-ph/0307037}
  {\path{arXiv:hep-ph/0307037}}, \href
  {http://dx.doi.org/10.1103/PhysRevD.68.094013}
  {\path{doi:10.1103/PhysRevD.68.094013}}.

\bibitem{Lappi:2016fmu}
T.~Lappi, H.~M{\"a}ntysaari, Next-to-leading order {Balitsky-Kovchegov}
  equation with resummation, Phys. Rev. D93 (2016) 094004.
\newblock \href {http://arxiv.org/abs/1601.06598} {\path{arXiv:1601.06598}},
  \href {http://dx.doi.org/10.1103/PhysRevD.93.094004}
  {\path{doi:10.1103/PhysRevD.93.094004}}.

\bibitem{Albacete:2018ruq}
J.~L. Albacete, G.~Giacalone, C.~Marquet, M.~Matas, Forward di-hadron
  back-to-back correlations in {$pA$} collisions from {rcBK} evolution{, }\href
  {http://arxiv.org/abs/1805.05711} {\path{arXiv:1805.05711}}.

\end{thebibliography}







\end{document}